\documentclass[]{spie}  

\def\bicep{\textsc{Bicep}}
\def\bicepone{\textsc{Bicep1}}
\def\biceptwo{\textsc{Bicep2}}
\def\bicepthree{\textsc{Bicep3}}
\def\biceptng{\textsc{Bicep} Array}
\def\keck{{\it Keck}}
\def\keckarray{{\it Keck Array}}
\def\sptpol{\textsc{SPTpol}}
\def\sptthreeg{\textsc{SPT-3G}}
\def\abs{\textsc{Abs}}

\def\class{\textsc{Class}}

\usepackage{amsmath,amsfonts,amssymb}
\usepackage{graphicx}
\usepackage[colorlinks=true, allcolors=blue]{hyperref}
\usepackage{url}
\usepackage{authblk}

\title{Ultra-Thin Large-Aperture Vacuum Windows for Millimeter Wavelengths Receivers}

\author[a]{Denis Barkats}
\author[a]{Marion I. Dierickx} 
\author[a,k]{John M. Kovac}
\author[b]{Chris Pentacoff}
\author[c]{P.~A.~R.~Ade}
\author[d]{Z.~Ahmed}
\author[e]{R.~W.~Aikin}
\author[a]{K.~D.~Alexander}
\author[g]{S.~J.~Benton}
\author[h]{C.~A.~Bischoff}
\author[e,i]{J.~J.~Bock}
\author[a]{R.~Bowens-Rubin}
\author[e]{J.~A.~Brevik}
\author[a]{I.~Buder}
\author[j]{E.~Bullock}
\author[a,k]{V.~Buza}
\author[a]{J.~Connors}
\author[a]{J.~Cornelison}
\author[i]{B.~P.~Crill}
\author[o]{M.~Crumrine}
\author[l]{L.~Duband}
\author[k]{C.~Dvorkin}
\author[m,n]{J.~P.~Filippini}
\author[o]{S.~Fliescher}
\author[f]{J.~Grayson}
\author[o]{G.~Hall}
\author[p]{M.~Halpern}
\author[a]{S.~Harrison}
\author[e,i]{S.~R.~Hildebrandt}
\author[q]{G.~C.~Hilton}
\author[e]{H.~Hui}
\author[f,d,p]{K.~D.~Irwin}
\author[f]{J. Kang}
\author[a,r]{K.~S.~Karkare}
\author[f]{E.~Karpel}
\author[s]{J.~P.~Kaufman}
\author[s]{B.~G.~Keating}
\author[e]{S.~Kefeli}
\author[f]{S.~A.~Kernasovskiy}
\author[f,d]{C.~L.~Kuo}
\author[o]{K.~Lau}
\author[r]{N.~A.~Larsen}
\author[r]{E.~M.~Leitch}
\author[e]{M.~Lueker}
\author[i]{K.~G.~Megerian}
\author[e]{L.~Moncelsi}
\author[t]{T.~Namikawa}
\author[i]{H.~T.~Nguyen}
\author[e,i]{R.~O'Brient}
\author[f,d]{R.~W.~Ogburn~IV}
\author[h]{S.~Palladino}
\author[j,o]{C.~Pryke}
\author[a]{B.~Racine}
\author[a]{S.~Richter}
\author[o]{R.~Schwarz}
\author[e]{A.~Schillaci}
\author[r,v]{C.~D.~Sheehy}
\author[e]{A.~Soliman}
\author[a]{T.~St.~Germaine}
\author[e,i]{Z.~K.~Staniszewski}
\author[e]{B.~Steinbach}
\author[c]{R.~V.~Sudiwala}
\author[e,s]{G.~P.~Teply}
\author[f,d]{K.~L.~Thompson}
\author[f]{J.~E.~Tolan}
\author[c]{C.~Tucker}
\author[i]{A.~D.~Turner}
\author[h]{C.~Umilt\`{a}}
\author[r,w]{A.~G.~Vieregg}
\author[f]{A.~Wandui}
\author[i]{A.~C.~Weber}
\author[p]{D.~V.~Wiebe}
\author[o]{J.~Willmert}
\author[a,k]{C.~L.~Wong}
\author[f,r]{W.~L.~K.~Wu}
\author[f]{H.~Yang}
\author[f,d]{K.~W.~Yoon}
\author[e]{C.~Zhang}

\affil[a]{Harvard-Smithsonian Center for Astrophysics, 60 Garden Street, Cambridge, MA 02138, USA}
\affil[b]{Massachusetts Institute of Technology, 77 Massachusetts Ave., Cambridge, MA 02139, USA}
\affil[c]{School of Physics and Astronomy, Cardiff University, Cardiff, CF24 3AA, United Kingdom}
\affil[d]{Kavli Institute for Particle Astrophysics and Cosmology, SLAC National Accelerator Laboratory, Menlo Park, CA 94025, USA}
\affil[e]{Department of Physics, California Institute of Technology, Pasadena, CA 91125, USA}
\affil[f]{Department of Physics, Stanford University, Stanford, CA 94305, USA}
\affil[g]{Department of Physics, University of Toronto, Toronto, Ontario, M5S 1A7, Canada}
\affil[h]{Department of Physics, University of Cincinnati, Cincinnati, OH 45221, USA}
\affil[i]{Jet Propulsion Laboratory, Pasadena, CA 91109, USA}
\affil[j]{Minnesota Institute for Astrophysics, University of Minnesota, Minneapolis, MN 55455, USA}
\affil[k]{Department of Physics, Harvard University, Cambridge, MA 02138, USA} 
\affil[l]{Service des Basses Temp\'{e}ratures, Commissariat \`{a} l'Energie Atomique, 38054 Grenoble, France}
\affil[m]{Department of Physics, University of Illinois at Urbana-Champaign, Urbana, IL 61801, USA}
\affil[n]{Department of Astronomy, University of Illinois at Urbana-Champaign, Urbana, IL 61801, USA}
\affil[o]{School of Physics and Astronomy, University of Minnesota, Minneapolis, MN 55455, USA}
\affil[p]{Department of Physics and Astronomy, University of British Columbia,Vancouver, British Columbia, V6T 1Z1, Canada} 
\affil[q]{National Institute of Standards and Technology, Boulder, CO 80305, USA} 
\affil[r]{Kavli Institute for Cosmological Physics, University of Chicago, Chicago, IL 60637, USA}
\affil[s]{Department of Physics, University of California at San Diego, La Jolla, CA 92093, USA}
\affil[t]{Leung Center for Cosmology and Particle Astrophysics, National Taiwan University, Taipei 10617, Taiwan} 
\affil[u]{Canadian Institute for Advanced Research, Toronto, Ontario, M5G 1Z8, Canada}
\affil[v]{Physics Department, Brookhaven National Laboratory, Upton, NY 11973}
\affil[w]{Department of Physics, Enrico Fermi Institute, University of Chicago, Chicago, IL 60637}

\authorinfo{Send correspondence to D. Barkats, 60 Garden Street, MS 42, Cambridge, MA 02138 
USA. E-mail: dbarkats@cfa.harvard.edu}

\begin{document} 
\maketitle

\begin{abstract}
Targeting faint polarization patterns arising from Primordial Gravitational Waves in the Cosmic Microwave Background 
requires excellent observational sensitivity. Optical elements in small aperture experiments such as \bicepthree~and \keckarray~are designed to 
optimize throughput and minimize losses from transmission, reflection and scattering at millimeter wavelengths. As aperture size increases, cryostat vacuum windows must withstand larger forces from atmospheric pressure and the solution has often led to a thicker window at the expense of larger transmission loss. We have identified a new candidate material for the fabrication of vacuum
windows: with a tensile strength two orders of magnitude larger than previously used materials,
woven high-modulus polyethylene could allow for dramatically thinner windows, and therefore significantly reduced
losses and higher sensitivity. In these proceedings we investigate the suitability of high-modulus polyethylene windows for ground-based CMB experiments, such as current and future receivers in the \bicep/\keckarray~program. This includes characterizing their optical transmission as well as their mechanical behavior under atmospheric pressure. We find that such ultra-thin materials are promising candidates to improve the performance of large-aperture instruments at millimeter wavelengths, and outline a plan for further tests ahead of a possible upcoming field deployment of such a science-grade window. \\
\end{abstract}

\keywords{Millimeter Wavelengths, Vacuum Windows, Polymer Materials, Cosmic Microwave Background, Primordial Gravitational Waves, Polarization, BICEP, Keck Array}

\section{INTRODUCTION}
\label{sec:intro}  

Many ground-based millimeter-wave receivers, such as those targeting the 2.7~K Cosmic Microwave Background (CMB), are designed around cryogenically cooled detectors and optics. Cryogenic receivers therefore include a vacuum window that separates cold components under vacuum from ambient air. Materials suitable for the fabrication of such windows must withstand the force developed by atmospheric pressure over the large area of the clear optical aperture, and minimize optical loading (transmission loss, reflections and scattering). These specifications lead to conflicting  solutions, as strong materials are typically not transparent in-band (Kevlar\textsuperscript{\textregistered}, Dacron\textsuperscript{\textregistered}, carbon fiber,  composites), while transparent materials have comparatively low bulk modulus and tensile strength (Polyethylene, Polypropylene, Teflon). CMB experiments so far have found a compromise by using various forms of polyethylene.

Zotefoam\textsuperscript{\textregistered}\footnote{http://www.zotefoams.com/} (HD30, PPA30), a Nitrogen-expanded polyethylene closed-cell foam, has been employed for experiments with aperture sizes smaller than $\sim13"$ (e.g. \bicepone~[\citenum{Yoon2006}], \biceptwo~[\citenum{Ogburn2014}], \keckarray~[\citenum{Vieregg2012}], SPT-SZ~[\citenum{Carlstrom2011}], \sptpol~[\citenum{SPTPol2012}], ACBAR~[\citenum{Runyan2003}], PolarBear~[\citenum{Kermish2012}]). Zotefoam\textsuperscript{\textregistered} has been used successfully in stacks up to 5'' thick for windows up to 13'' in diameter. With in-band transmission exceeding $99$\% and low index of refraction $n\sim1$, no anti-reflective (AR) coating is necessary. This made Zotefoam\textsuperscript{\textregistered} an ideal candidate for mm-wave windows. However, scaling to larger diameter and thicker layers has proved too cumbersome.

For larger apertures, where the required thickness of Zotefoam\textsuperscript{\textregistered} would become unwieldy, slabs of bulk high-density  polyethylene or ultra-high molecular weight polyethylene (HDPE and UHMWPE) have been the default solution (e.g. for \bicepthree~[\citenum{Ahmed2014, Grayson2016}], \abs~[\citenum{Essinger-Hileman2010}], \sptthreeg~[\citenum{Benson2014}], \class~[\citenum{Essinger-Hileman2014}], QUIET~[\citenum{Bischoff2013}], ACT-MBAC~[\citenum{Swetz2011}], ACTPol and AdvACT~[\citenum{Thornton2016}]). The transmission properties of these materials have been studied in the lab (e.g.~[\citenum{Dalessandro2018}]) and their loss is known to scale with the thickness. With an index $n=1.53$, antireflection-coating becomes necessary to minimize reflection losses.

   \begin{figure} [h!]
   \begin{center}
   \begin{tabular}{c} 
   \includegraphics[height=6.2cm,trim=7mm 0mm 15mm 0mm,clip]{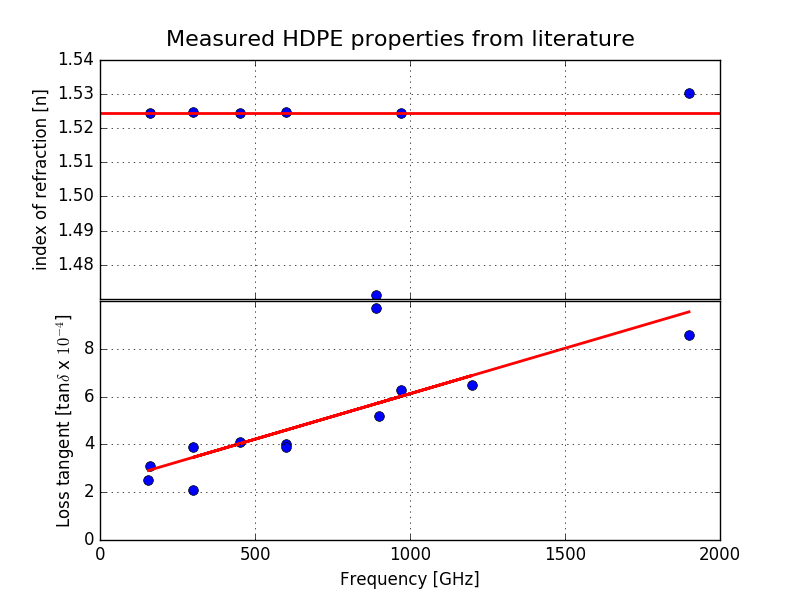} \\
   \includegraphics[height=6.2cm, trim = 5mm 0mm 16mm 12mm, clip]{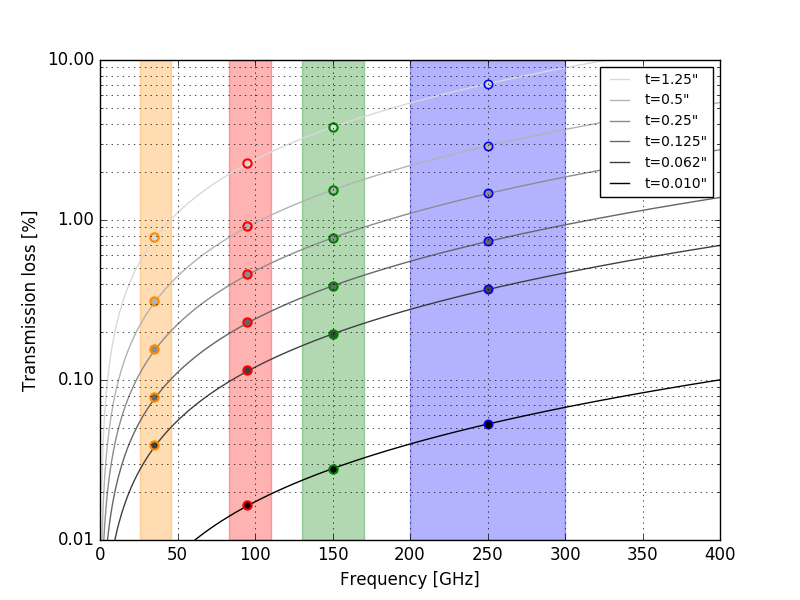}
   \includegraphics[height=6.2cm, trim = 10mm 0mm 0mm 12mm, clip]{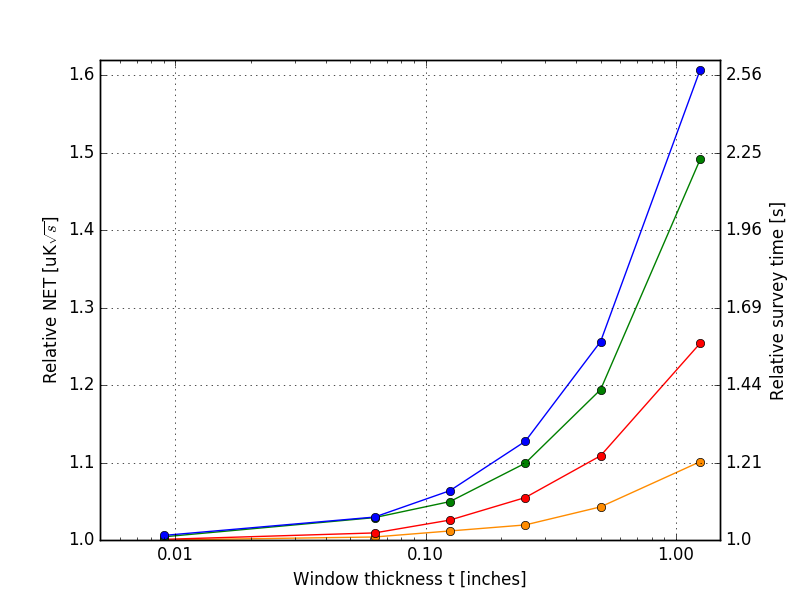}
   \end{tabular}
   \end{center}
   \caption[example] 
   { \label{fig:NET} 
     \textbf{Top:} Measured HDPE index of refraction and loss tangent from multiple references in~[\citenum{Lamb1996}]. In subsequent figures and calculations, we assume HDPE $n=1.53$ and $\tan\delta = 2\!\times\!10^{-4}$ for frequencies below 300~GHz. \textbf{Bottom Left:} Transmission loss for HDPE windows of 6 different thicknesses (ranging from 0.01" to 1.25") as a function of frequency, calculated assuming a loss tangent for HDPE of $2\!\times\!10^{-4}$. The standard atmospheric observing bands are shown in orange (30-40~GHz), red (100~GHz), green (150~GHz), and blue (220-270~GHz). \textbf{Bottom Right:} Relative Noise-Equivalent Temperature (NET, in $\mu$K$\sqrt{\text{s}}$) or, relative survey time (in seconds), as a function of window thickness, for different observing frequencies corresponding to the color-coded points in the left panel. This figure is generated using the estimated transmission loss from points in the middle panel as input in an NET calculator, assuming all other instrument parameters remain fixed to BICEP3 [\citenum{Ahmed2014,Grayson2016}] values. The NET and survey time shown are relative to an ideal no-window case normalized to 1. The degradation in sensitivity introduced by the window is worse in higher frequency bands. For example, a 1/2"-thick window at 150~GHz would increase the NET by 20\%, or would require a 44\% longer survey time to reach the same noise level.
}
   \end{figure}

To achieve ever-increasing  sensitivities, CMB experiments depend on a steady increase in optical throughput (more detectors in the focal plane and larger focal planes),  which naturally translates into larger optical apertures. The thickness of HDPE/UHMWPE slabs used as vacuum windows must increase accordingly in order to withstand larger forces from atmospheric pressure. Figure~\ref{fig:NET} shows the dependence of transmission loss and relative sensitivity on nominal window thickness, for different observing frequencies assuming the standard bulk polyethylene windows. The Noise-Equivalent-Temperature (NET) is a steep function of window thickness, especially at higher frequencies. Above 100~GHz, even a modest thickness of HDPE significantly affects the sensitivity on the sky, as the window begins to  contribute significantly to the total optical loading on the detectors. `Stage 3' CMB experiments  are planning receivers with apertures in excess of 60~cm (for example \biceptng's 28" diameter receivers) with several bandpasses up to 270~GHz. Improvements in vacuum window design and fabrication are necessary in order to prevent degradation in sensitivity as observations push to higher frequencies. 

\begin{figure} [bt]
   \begin{center}
   \begin{tabular}{c} 
   \includegraphics[height=5.1cm, trim = 0mm 0mm 0mm 0mm, clip]{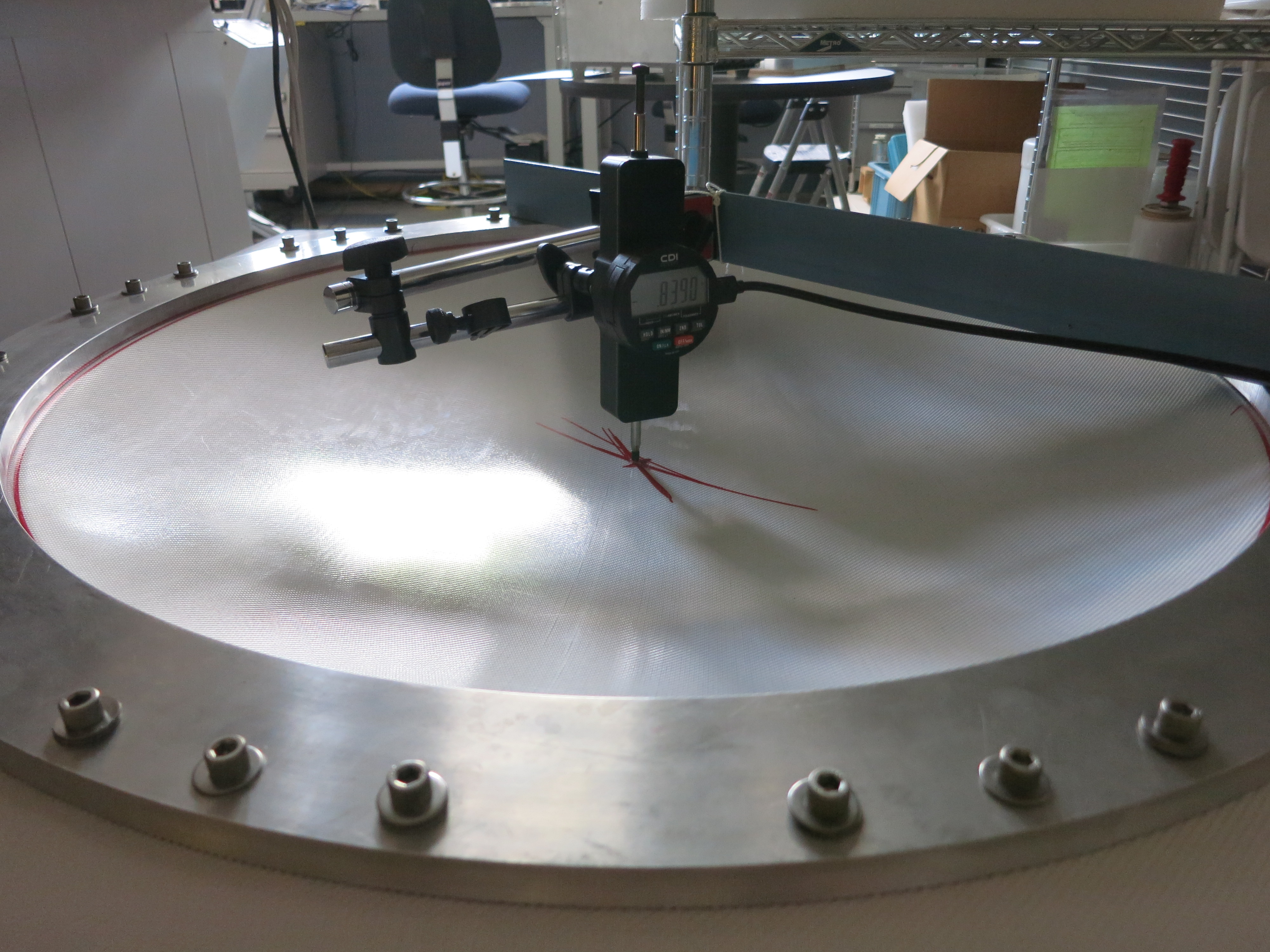}
      \includegraphics[height=5.1cm]{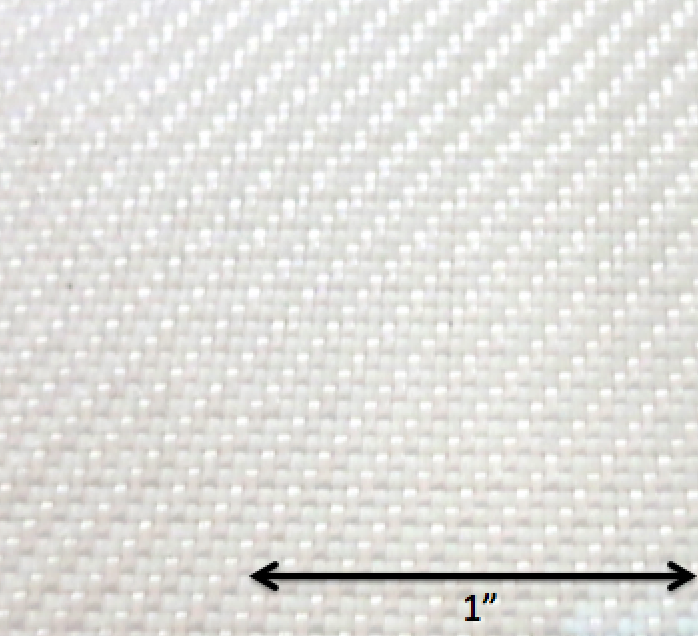}
       \includegraphics[height=5cm, trim = 4mm 0mm 0mm 0mm, clip]{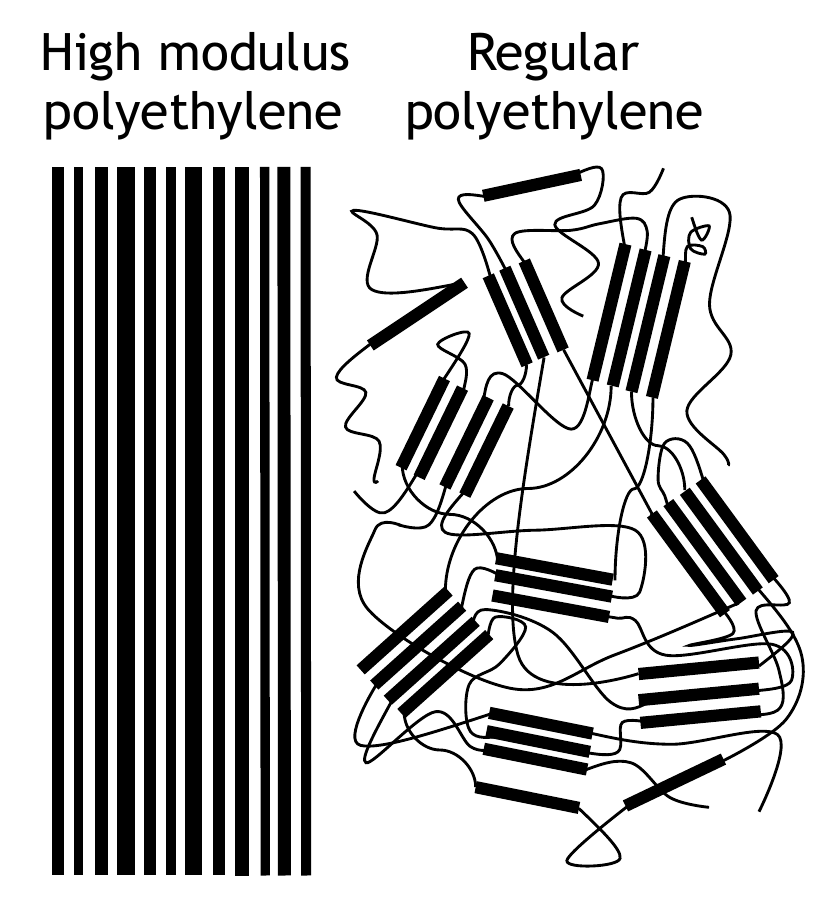}
   \end{tabular}
   \end{center}
   \caption[] 
   { \label{fig:dyneema_photos} 
\textbf{Left:} photos of a prototype HMPE vacuum window with 27'' clear aperture under test on a vacuum chamber. A depth gauge measures the deflection at the center of the window as a function of time. \textbf{Middle:} Closeup photo showing the texture of the woven fabric. \textbf{Right:} Cartoon representation of molecular alignment obtained in high-modulus polyethylene compared to the bulk polyethylene.} 
   \end{figure}

Since the 1990s, advancements in gel spinning technology have allowed the commercial development of spun UHMWPE fibers known as Dyneema\textsuperscript{\textregistered}\footnote{Dyneema\textsuperscript{\textregistered} is a regisitered trademark of Royal DSM N.V.} or Spectra\textsuperscript{\textregistered}\footnote{Spectra\textsuperscript{\textregistered} is a regisitered trademark of Honeywell International.}. For the rest of this paper, we refer to these fibers as High Modulus Polyethylene (HMPE). Extruding UHMWPE gel through a spinneret at a carefully controlled temperature leads to a high degree of molecular chain alignment (see Figure~\ref{fig:dyneema_photos}). The alignment of the fibers, or crystallinity, yields a quoted tensile modulus and tensile strength approximately 100 times higher than that of the bulk UHMWPE. The tensile strength of HDPE is typically between 20 and 40 MPa  while that of HMPE is 1-4 GPa [\citenum{dsm1,dsm2,dsm3,dsm4}].  Thin, fabric-like sheets can be woven from individual fibers and coated to prevent fraying. With a strength-to-weight ratio exceeding that of steel by a factor of approximately 10, industrial applications of HMPE have included radomes, ballistics protection, lightweight link chains and a wide range of technical fabrics.

Since these HMPE fibers are fabricated from Ultra High Molecular Weight Polyethylene, we expect they might have similar refractive index and loss tangent at millimeter wavelengths as the bulk material. The combination of increased strength and likely suitable optical properties presents the potential of dramatically thinner (and therefore more transparent) vacuum windows. Here we present initial results of an investigation aiming to test the relevant mechanical and optical characteristics of windows made of woven HMPE fabric. An example prototype with 27'' clear aperture is shown under test on a vacuum chamber in Figure~\ref{fig:dyneema_photos}. The woven-like nature of the `raw' HMPE fabric requires further custom processes to laminate it with other plastics to obtain a vacuum tight surface, to prevent fraying, and to transform the woven matrix into an optically homogeneous material.

The outline of these proceedings is as follows. In Section~\ref{sec:mechanical} we present tests aiming to validate the high tensile characteristics of the HMPE fibers and to investigate the behavior of prototype windows placed under atmospheric pressure for long periods of time. We have experimented with various fabrication processes and are therefore including results for different samples and prototypes. Throughout these proceedings we use the term `raw Dyneema\textsuperscript{\textregistered}' or `raw HMPE' to refer to the unlaminated woven fabric of high modulus fibers as received from the manufacturer. The samples referred to as `single layer' or `double layers' are made of one or more layers of raw HMPE fabric laminated in between thin layers of LDPE on either side. We are also experimenting with a different form of non-woven HMPE fabric, which we refer to as `CT10', consisting of wider flat-pressed sheets of HMPE laminated together. We note that we are still optimizing the window fabrication process and do not discuss the details here.
In Section~\ref{sec:optical} we present results from reflection spectra of HMPE samples that confirm their index of refraction is consistent with that of bulk UHMWPE. Section~\ref{sec:future} describes plans for future work paving the way to a full technology demonstration and eventual field deployment of a large-aperture HMPE window on BICEP3 at the South Pole. Section~\ref{sec:conclusions} summarizes our Conclusions.

\section{Mechanical Properties of HMPE  Window Prototypes}
\label{sec:mechanical} 

   \begin{figure} [hb]
   \begin{center}
   \begin{tabular}{c} 
   \includegraphics[height=9cm, trim = 0mm 0mm 0mm 10mm, clip]{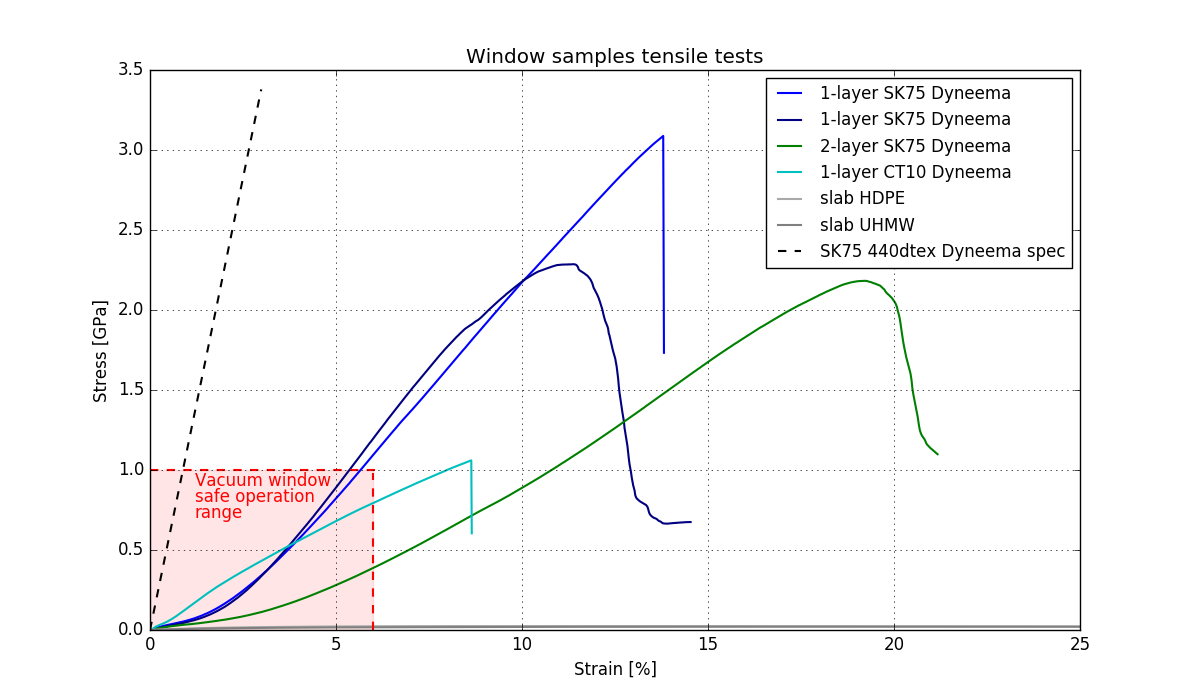}
   \end{tabular}
   \end{center}
   \caption[example] 
   { \label{fig:instron} 
Measured stress vs. strain curves for various samples of woven HMPE fabrics. The dashed line shows the manufacturer-quoted specification for raw Dyneema\textsuperscript{\textregistered} yarn, with a tensile modulus of 113~GPa, a tensile strength at yield of 3.4~GPa and yield strain of 3.5\%. The two blue lines both correspond to single-layer HMPE  samples and are included to illustrate different failure mechanisms. While one sample (blue) exhibits a sharp transition at yield, corresponding to a sudden ripping in the bulk of the material, the other (navy) fails  gradually due to stress concentration at the gripping zone. The two-layer sample shown in the green line exhibits a similar slow failure and thus does not probe the maximum stress the sample could withstand. The red box illustrates a suggested zone of safe operating conditions (the values chosen here below 1~GPa stress and 6\% strain) where the materials remain in elastic deformation regime far from their failure point. Note the gray lines near the bottom of the plot showing measurement curves  of reference HDPE and UHMWPE samples.}
   \end{figure} 

The first step in assessing the suitability of woven HMPE fabric for vacuum windows is to understand its behavior when placed under mechanical stress. 
We carried out different types of tests, with the aim to: 
\begin{itemize}
\item confirm the manufacturer-quoted tensile properties of these HMPE fibers,
\item compare it to previously used materials (bulk HDPE/UHMWPE),
\item and assess the long-term behavior of realistic science-grade windows under realistic operating conditions. 
\end{itemize}

As a first step, we measure the stress-strain relationship of different HMPE window variants and compare them to manufacturer specification and to reference bulk HDPE/UHMWPE samples. We obtain curves for force (in Newtons) vs. extension (in mm) using an Instron\textsuperscript{\textregistered}\footnote{http://www.instron.us/} testing system. Those curves are then converted to stress (in units of Pascals) vs. strain (in \%) based on the measured lengths and cross-sectional area of our samples. We note that it is critical to carefully account for the correct cross-sectional area of the fabric based on its matrix structure, not on its bulk properties treating it as a homogeneous material. Figure~\ref{fig:instron} presents representative profiles for our different types of samples. The resulting modulus, yield stress and yield strain values are summarized in Table~\ref{tab:instron}.  We find that the yield strength (the maximum stress on the curves just before failure) for our laminated single-layer sample closely matches the manufacturer-quoted specification for the HMPE yarn. We also find that all our laminated samples are more elastic (lower Young's modulus and higher strain at yield) than the specification for the single yarn. This result is not unexpected. Compared to a individual yarns, the woven structure of the HMPE fabric tends to straighten itself under tension, yielding more elastic properties than the individual yarn. We also tested 2-layer laminates samples. We confirmed that the holding force scales with cross section (yield strength in Pa remained equal for 1- and 2-layer laminates) but noted an unexpected increase in elasticity. We suspect this is caused by imperfect bonding between the two woven fabric layers and are still exploring the optimal bonding procedure to reduce this effect. In the end, all our tested HMPE woven laminates were significantly stronger and stiffer than the bulk HDPE/UHMWPE (whose stress vs. strain curves are barely visible near the bottom of Figure~\ref{fig:instron}). For vacuum window applications, we plan to fabricate windows that remain well within the elastic deformation regime, approximately below 1~GPa stress and 6\% strain for the samples tested.

\begin{table}[t]
\label{tab:instron}
\begin{center}       
\begin{tabular}{|l|l|l|l|} 
\hline
\rule[-1ex]{0pt}{3.5ex}  Material & Tensile modulus [GPa] & Yield stress [GPa] & Strain at yield [\%]  \\
\hline
\rule[-1ex]{0pt}{3.5ex}  Dyneema\textsuperscript{\textregistered} spec & 113 & 3.4 & 3.5  \\
\hline
\rule[-1ex]{0pt}{3.5ex} 1-layer woven HMPE fabric & 26.9 & 3.1 & 11.9   \\
\hline
\rule[-1ex]{0pt}{3.5ex} 2-layer woven HMPE fabric & 14.3 & 2.2 & 16.4  \\
\hline
\rule[-1ex]{0pt}{3.5ex} 1-layer CT10 fabric  & 11.9 & 1.1 & 9.3   \\
\hline
\rule[-1ex]{0pt}{3.5ex} Bulk HDPE & 0.9 & 0.025 & 10  \\
\hline
\rule[-1ex]{0pt}{3.5ex} Bulk UHMWPE & 0.7 & 0.022 & 30  \\
\hline 
\end{tabular}
\end{center}
\caption{Material properties measured from structural tests for different samples in an Instron\textsuperscript{\textregistered} structural testing instrument. In order to account for the varying thicknesses of the samples and to make the data directly comparable, we have scaled the force by the cross-sectional area of each sample and provided the results in units of stress (in GPa) and strain (in \%).
}
\end{table}

\begin{figure} [t]
\begin{center}
\begin{tabular}{c} 
\includegraphics[height=8.5cm, trim = 0mm 0mm 0mm 15mm, clip]{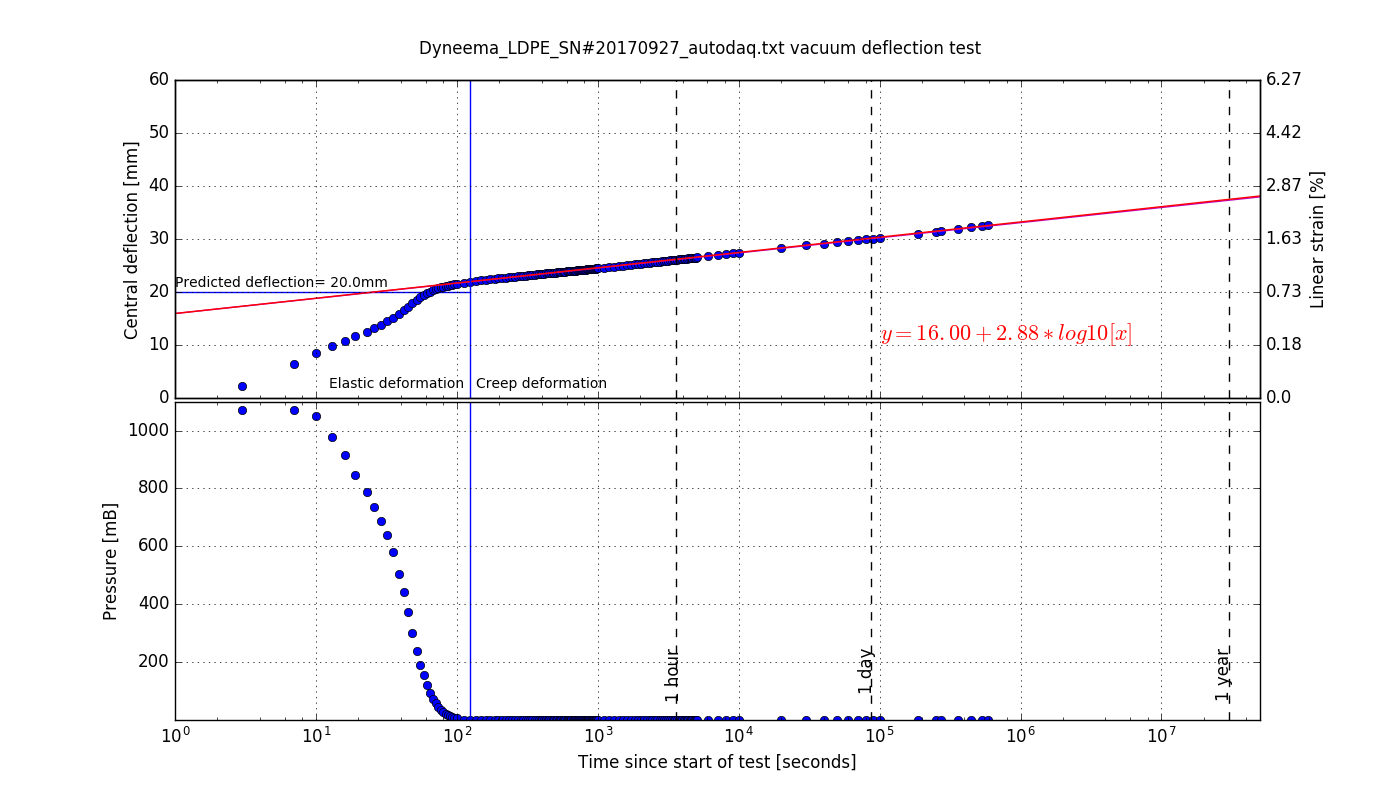}
\end{tabular}
\end{center}
\caption[example] 
{ \label{fig:deflection} Measured deflection (\textbf{top panel}) and pressure (\textbf{bottom panel}) as a function of time for a 15"-diameter single-layer woven HMPE window tested on a vacuum chamber for 3 months. After the initial elastic deflection associated with the rapid pressure drop, the window remains under constant load (2600 lbf = 11600 N) and experiences a continuous creep that results in increased deflection over time. The red line is a fit to a logarithmic creep rate with time. Under these conditions of pressure and temperature, we expect the strain to remain under 3\% after 1-year of constant load, well within the safe operating regime  shown in Figure~\ref{fig:instron}. }
\end{figure} 

\begin{figure} [t]
\begin{center}
  \begin{tabular}{c} 
    \includegraphics[height=6.1cm, trim = 15mm 5mm 25mm 10mm, clip]{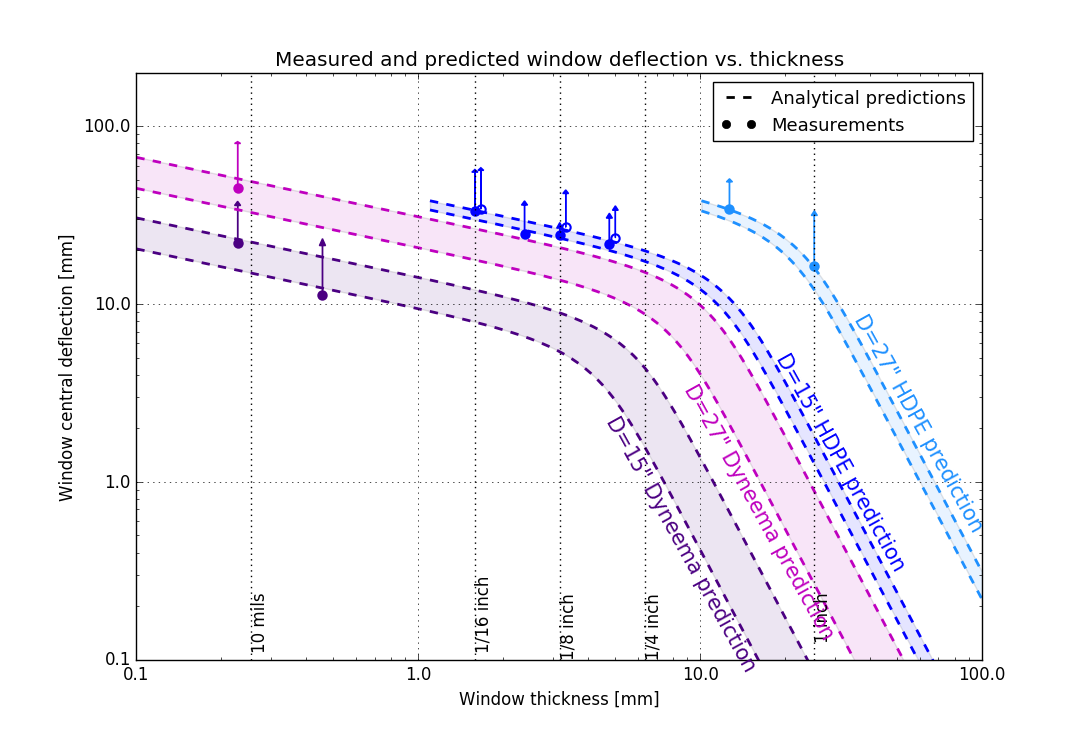}
    \includegraphics[height=6.1cm, trim = 18mm 5mm 0mm 10mm, clip]{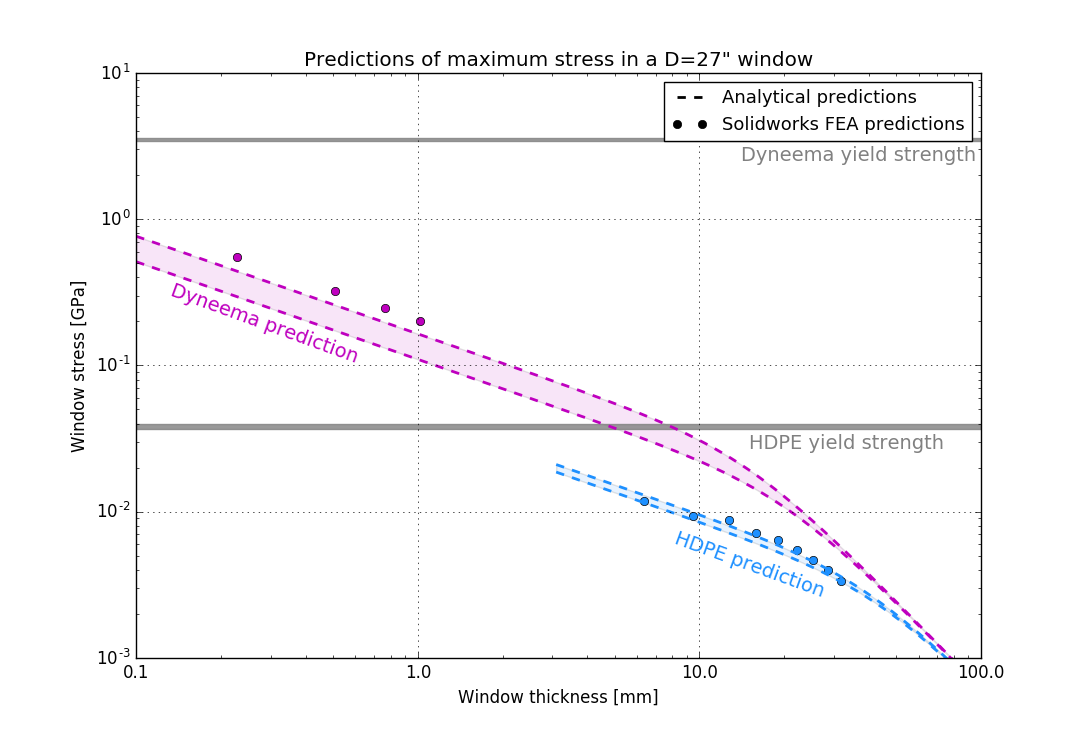}
  \end{tabular}
\end{center}
\caption[example] 
{ \label{fig:deflection_summary} 
\textbf{Left:} Measured (dots) and analytic predictions (colored bands) of window deflection as a function of thickness. The plot shows calculations for woven HMPE fabrics (purple) and standard bulk HDPE (blue) for D=15" and D=27" windows. The predictions are truncated at deflections corresponding to a safety margin of 2 (stress = 1/2 yield stress). The width of the bands reflects the uncertainty in the materials' Young's moduli. Arrows above each point show the extrapolated creep after 1 year based on a logarithmic fit (example shown in Figure~\ref{fig:deflection}). For the D=15'' HDPE windows, we have measured the deflections using both HDPE windows (solid blue points) and UHMWPE windows (open blue points) and found no significant difference in mechanical behavior in those tests. 
\textbf{Right:} FEA calculations (dots) and analytic predictions (colored bands) of window stress as a function of thickness, for the full-size D=27" windows only. The analytical predictions are truncated at 50\% of the nominal yield stress. For the FEA simulations, the values shown are the initial median stress experienced in each simulation. No modelling of creep is included. There is good agreement between the stress profiles derived from the two methods. Grey lines show the nominal maximum stress sustainable by both standard HDPE (40~MPa) and HMPE (3.5~GPa).}
\end{figure}

Moving beyond unit-level verification of material properties, we seek to assess the long term creep (visco-elastic deformation) of realistic science-grade windows under operating conditions. We have at our disposal two test vacuum chamber with 15'' and 27'' diameter aperture. The 27''-diameter chamber nearly matches the diameter of \bicepthree~and planned \biceptng~receiver windows and therefore provides an ideal testing setup for those future windows. Figure~\ref{fig:deflection} shows the results of a long duration test where a prototype single-layer woven HMPE, 15"-diameter window was placed under vacuum for several months. Initially the window deflects suddenly as a result of the rapid pressure drop. Once the pressure inside the chamber stabilizes to a low value ($P<0.1$~mBar), the window experiences the maximum force from atmospheric pressure and continues to deform logarithmically in time. In this creep deformation regime, we can extrapolate the deflection to a full year and show it remains well within the safe operating zone outlined in Figure~\ref{fig:instron} (strain~$\lesssim 5\%$). We have tested many windows, including thicker bulk HDPE and thin laminated woven HMPE fabrics, for durations ranging from 1 week to 6 months, on  our 15'' and 27''-diameter test chambers  and the results are summarized in Figure~\ref{fig:deflection_summary}.

Figure~\ref{fig:deflection_summary} (left) shows the measured initial deflections (solid circles), the extrapolated deflection after 1 year (arrows), and the analytical deflection predictions (dashed bands). The agreement between the measured initial deflection and the simple analytical expectation is excellent. The deflection due to creep is not accounted for in the calculation. The analytical predictions for deflection follow the model given by Equation~\ref{eq:deflection1} from [\citenum{Timoshenko}]:
\begin{equation}\label{eq:deflection1}
\delta + 0.488\;\frac{\delta^3}{t^2} = \frac{3}{16}(1-\nu^2)\frac{{dP}\;R^4}{E\;t^3}
\end{equation}
where $\delta$, $t$, and $R$ are the window material central deflection, thickness, and radius in meters, $\nu$ is the material's unitless  Poisson's ratio, $dP$ is the pressure differential on the window, and $E$ is the material's (flexural or tensile) modulus. We can see that in the linear regime of small deflection to thickness ratio (bottom right of Figure~\ref{fig:deflection_summary} left panel), the deflection scales as 
\begin{equation}
\delta \propto \frac{dP\;R^4}{E\;t^3}
\end{equation}
 At the other end of the spectrum, in the regime of large deflection to thickness ratio (top left of Figure~\ref{fig:deflection_summary} left panel), the deflection scales as
\begin{equation}
\delta \propto \left(\frac{dP\;R^4}{E\;t}\right)^{\frac{1}{3}}
\end{equation}
The windows made of ultra-thin HMPE woven laminates presented here are taking advantage of this latter regime where the deflection increases more slowly with thickness. With a 27''-diameter aperture, we have indeed obtained similar deflections with 0.5''-thick HDPE ($\delta=34$~mm) as with a 0.010''-thick woven HMPE laminated window ($\delta=45$~mm).

We also compute analytical predictions for the maximum stress experienced by these windows. The equations for stress are also taken from [\citenum{Timoshenko}] and can be written as simple expressions in the two limiting cases. In the regime with large deflection to thickness ratio, we have
\begin{equation}\label{eq:stress1}
\sigma = K \left(\frac{E\;dP^2\;R^2}{t^2}\right)^{1/3}
\end{equation}
and in the regime of large deflection to thickness, we obtain
\begin{equation}\label{eq:stress2}
\sigma = K \left(\frac{(1+v)\;dP\;R^2}{t^2}\right)
\end{equation}
where $K$ is a unitless factor between 0 and 1, and $\sigma$ is the maximum stress on the window.
The stress predictions for a nominal 27''-diameter window are shown in the right panel of Figure~\ref{fig:deflection_summary} for both bulk HDPE and HMPE. The colors correspond to those used for the deflection curves on the left side of Figure~\ref{fig:deflection_summary}. The analytical curves are truncated at a nominal safety factor of two (where the stress reaches half the nominal yield strength of the material). We have taken the yield strength of HDPE to be 40~MPa and that of HMPE to be 3.5~GPa. To form complete curves, we have combined the expressions for the two limiting cases in inverse quadrature. We validate the analytical predictions of stress with accompanying Finite Element Analysis (FEA) simulations and find good agreement in the overall shape. These predictions indicate that for 27''-diameter, a 0.010''- to 0.050''-thick HMPE window will nominally provide a larger safety factor than $1/4$'' to $1/2$''-thick HDPE/UHMWPE window. We also note that these simulations and deflection tests were all performed  with sea level atmospheric pressure of 14.7~psi (1013~mBar). Actual deployment on an instrument at the South Pole will have significantly less mechanical loading. (The South Pole atmospheric pressure is 9.7~psi or 670~mBar.)

\section{Optical Properties of HMPE Samples}
\label{sec:optical}  

    \begin{figure} [ht]
   \begin{center}
   \begin{tabular}{c} 
   \includegraphics[height=7cm, trim = 0mm 5mm 0mm 10mm, clip]{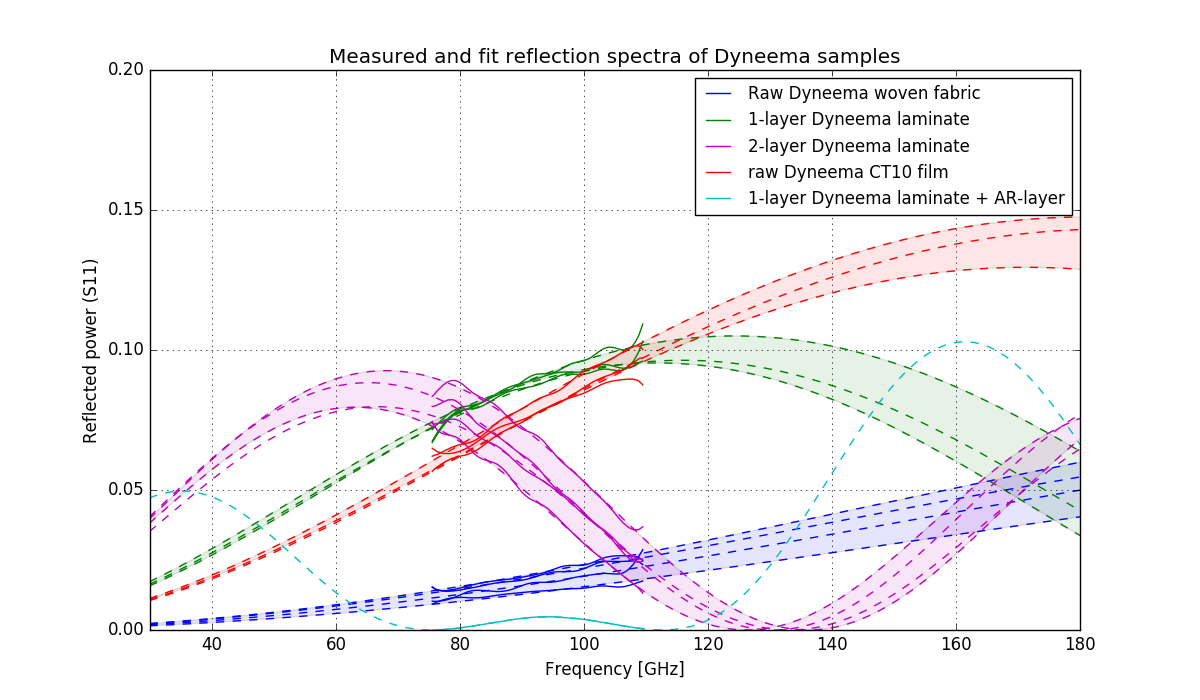} \\
      \includegraphics[height=7cm, trim = 0mm 5mm 0mm 10mm, clip]{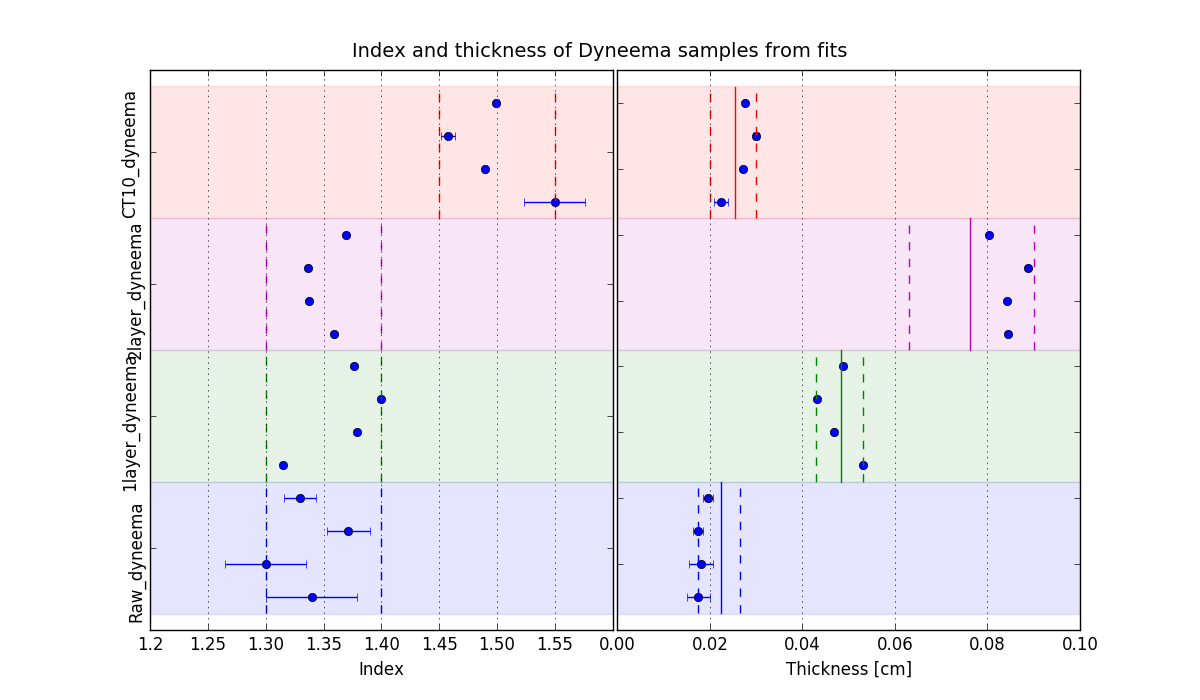}
   \end{tabular}
   \end{center}
   \caption[example] 
   { \label{fig:reflection} 
\textbf{Top Panel:} Measured Reflected power (S11) spectra of various samples of woven HMPE fabric at 75-110~GHz (solid lines). For each type of fabric, we repeat the measurement 3-4 times with different samples from the same batch to constrain measurement uncertainties. We fit those spectra to a multi-layer reflectance model (dashed lines) to extract index and thickness. The fit parameters closely match the measured thickness and the expected index of refraction of UHMWPE ($n\sim1.53$). For the woven HMPE fabrics, we find that a $\sim40-60$\% fill factor compared to a homogeneous material explains the lower index of refraction ($n\sim1.35$). The cyan line shows the low reflected power from an AR-coated one-layer HMPE fabric sample. 
\textbf{Bottom Panel:} Index of refraction and thickness recovered from fits to data shown in top panel (with matching colors). Vertical dashed lines show the fit boundaries. The vertical solid line is the measured thickness of the samples. The recovered index of refraction of the woven HMPE fabric ($1.3<n<1.4$) is consistently lower than that of bulk UHMWPE ($n\sim1.53$) because the woven fabric has a lower density even after lamination.
}
   \end{figure} 

Ensuring the mechanical resilience of vacuum windows is critical for overall system safety. For HMPE windows, characterizing the expected optical transmission loss and the possible scattering associated with these new materials is also essential. Optimal millimeter-wave vacuum windows must have minimal bulk loss as well as low reflectivity and scattering. Reflected light can be minimized through the application of an AR coating layer, with $\lambda/4$ thickness and an index of refraction approximately equal to the square root of that of the bulk material. For example, layers of expanded Teflon ($1.18< n <1.24$) on either side of an UHMWPE ($n\sim1.53$) window can reduce the band-averaged reflections to less than 1\%. Since the HMPE fabric windows we are testing are made from woven fibers of UHMWPE, we hypothesize that their index of refraction and transmissivity may be similar to a thin layer of bulk UHMWPE.  Here we present reflection tests carried out with the aim of:
\begin{itemize}
\item confirming high transmission consistent with that of UHMWPE, and
\item demonstrating we can minimize reflectivity through the application of an AR coat.
\end{itemize}

We use a Vector Network Analyzer (VNA) to produce a frequency sweep from 75 to 110~GHz and measure the reflected spectra (S11) from  various HMPE window samples. The signal is coupled to free space with a standard gain horn, and the beam is then collimated with a $90^\text{o}$ refocusing mirror. The samples are placed in the near-field of the collimating optics to obtain nearly-orthogonal angles of incidence on the sample. The samples are carefully tensioned in a stretching device in order to produce a flat, repeatable surface of incidence. In addition to the standard in-waveguide  short/load calibration, we apply a time-domain gating filter to the measured S11 response to isolate the free space reflections from our samples. For the raw, one-layer and two-layer samples, data is acquired multiple times from distinct samples from the same batch, in order to evaluate measurement error and sample-to-sample variations. The resulting spectra are shown in Figure~\ref{fig:reflection}. We also include the reflection spectrum measured for a single-layer sample laminated with an expanded Teflon (Teadit\textsuperscript{\textregistered}) AR coating layer. The data are fit with a theoretical spectrum for a stratified medium computed with the transfer matrix method.

From these fits, we recover the thickness and index of refraction for each sample and show the results in Figure~\ref{fig:reflection}. All three woven HMPE fabric samples (raw, single- and double-layer) present an index $n\sim1.35$, somewhat lower than the expectation for bulk UHMWPE ($n\sim1.53$). This low index is indeed consistent with the density of the woven HMPE fabric, whether laminated or not. Measurements of the density suggest a 40-60\% fill factor. This interpretation is reasonable given the interweaving of fibers in the fabric is expected to give rise to voids. We expect the recovered density and index of refraction to strongly depend on the lamination process parameters (temperature, pressure, and time) and this dependence is still under investigation.  Altogether, our reflection measurements confirm that the reflection characteristics of woven HMPE fabric are similar to those of the bulk UHMWPE material. We expect the (more difficult to measure) transmission loss to yield similar results. We also demonstrate that the in-band reflected power can be strongly reduced through the application of a standard AR layer. One additional advantage of these ultra-thin laminated windows is that their intrinsic thickness can be tuned ahead of AR coating to minimize reflections around a particular frequency. In combination with standard AR layers, this can produce low reflections over very wide bandwidths. We eventually plan to extend these free space reflectivity tests down to 30~GHz and up to 270~GHz to cover the expected bandpasses of \biceptng~receivers.

\section{Future Work and Pathway to Field Deployment}
\label{sec:future}  

Further characterization of both the mechanical behavior and optical properties of these woven HMPE windows is needed prior to field deployment. Beyond the results presented in these proceedings, here we outline ongoing and future tests to be carried out before integration into a working receiver. 

\subsection{Testing and Optimization of Mechanical Properties}

The next step is to scale up the testing effort in order to replicate the full \bicepthree~and \biceptng~  aperture size and flange design in the lab. In parallel, we plan to carry out long duration tests in order to study the creep and deflection behavior over longer ($>6$ months) timescales than those shown in Figure~\ref{fig:deflection}. Another area of interest for the windows featuring multiple layers of HMPE is the effect of fiber alignment. Because of the woven matrix structure of these fabrics, they mainly withstand stress in two orthogonal directions, whereas a window under load develops forces in all radial directions. We are therefore in the process of investigating the advantages and drawbacks of laminating multiple layers at different angles from each other, compared to co-aligning all layers. Preliminary tests suggest that co-alignment of the fibers allows for higher yield stresses, likely because it helps maintain layer cohesion. However, these results must be confirmed once the lamination process parameters are optimized. Further validations needed prior to field deployment include investigating cold temperature creep behavior, leak rates, UV-light stability, and failure modes during destructive testing. While many of these properties are already known and largely satisfactory for the HMPE fiber itself [\citenum{dsm3}], we nevertheless plan further tests of the most important properties to our application. We also note that the combination of lower atmospheric pressure at the South Pole and the lower operating temperature of the windows are both expected to significantly decrease HMPE fiber creep rate, possibly by 1-2 orders of magnitude~[\citenum{dsm0}] compared to lab conditions, and therefore contribute to increasing the lifetime of the window.

\begin{figure} [ht]
   \begin{center}
   \begin{tabular}{c} 
   \includegraphics[height=6cm, trim = 20mm 0mm 28mm 0mm, clip]{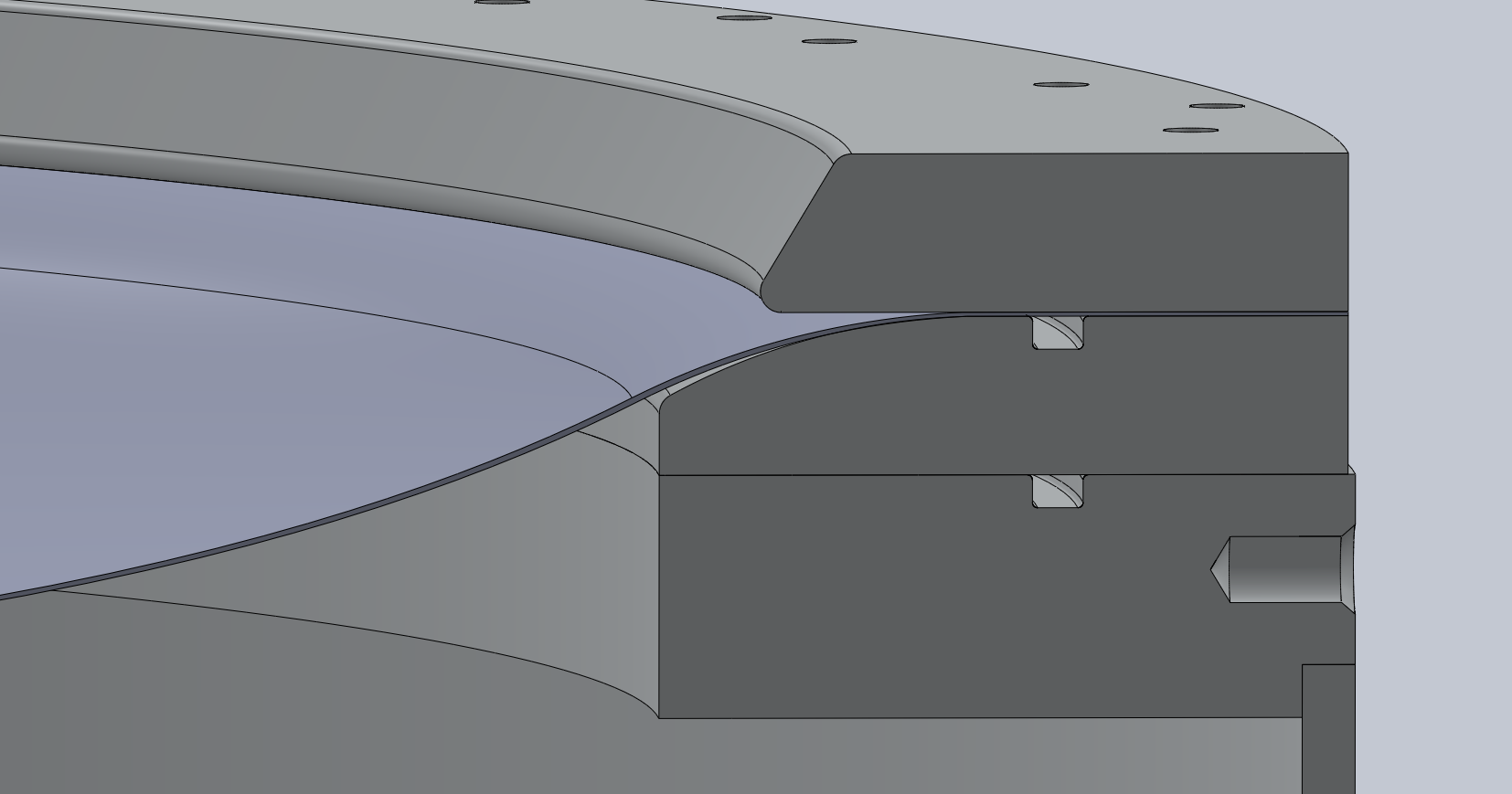}
   \end{tabular}
   \end{center}
   \caption[example] 
   { \label{fig:window_flange} 
Cutaway initial design of the thin window frame planned for the \biceptng~receiver.

}
   \end{figure}

In a parallel effort, we are working to optimize the design of the vacuum window flanges. The large force of atmospheric pressure directed downward on the window can cause it to slip inwards and  accumulate high stresses around the  clamping points. As a result, we are experimenting with special  clamping features in order to improve gripping forces on the plastic surface, for example using different designs of knurling and serrations. 
Another aspect of frame design we are currently examining is the transition from supported to unsupported surfaces around the window edge. At those locations, the windows  experience a steep sudden deflection that can cause stress accumulation. A smoother  transition with optimal curvature to match the natural launch angle of the deflected window will contribute to reduced shear in thicker materials, and reduced  stress concentrations around the edge. Figure~\ref{fig:window_flange} shows an example design of window frame and clamp where the edge of the HMPE  window under load conforms to a curved flange. 

\subsection{Optical Characterization} 
Along with loss due to absorption and reflection, scattering from optical elements is detrimental as it increases loading  and can introduce systematic noise in CMB observations if the scattered light is not properly baffled. The woven, mesh-like nature of these HMPE windows could conceivably introduce additional scattering compared to homogeneous UHWMPE. We are investigating this possibility using two complementary approaches. During the 2017-2018 austral summer, data was gathered with prototype raw, 1-layer, 2-layer HMPE windows placed in front of the \bicepthree~and \keckarray~receivers at 95, 210 and 270 GHz. The procedure included datasets  which should enable us to constrain the integrated scattered radiation. Secondly, we plan to measure scattered spectra at different angles away from standard specular reflection in the lab with a VNA. We specifically wish to test the hypothesis that samples laminated and filled in with LDPE will appear optically more homogeneous to radiation and therefore exhibit weaker scattering than the raw material. Finally, the crystalline woven nature of the HMPE  fabric may introduce polarization effects. We are planning to characterize the polarization-dependent index of refraction of the HMPE windows and to confirm that it is not birefringent on the  large scales probed by the beams of the \bicep/\keck~telescopes.

\section{Conclusions}
\label{sec:conclusions}  

For experiments with photon-limited detectors, probing the  faint polarization patterns in the CMB requires ever-increasing sensitivity and therefore expanding focal plane arrays. The aperture size of CMB telescopes has been growing, driving the need for new technology development. In particular, future large-aperture receivers will have to meet the challenge of fabricating larger optical elements that retain structural integrity without introducing prohibitively large losses.
Ultra-thin windows made from woven high modulus fibers of UHMWPE are promising candidates for large-aperture receivers operating at millimeter wavelengths. By providing an order of magnitude reduction in loss from the window, they could lead to a significant reduction in the photon loading from the instrument and therefore an improvement in sensitivity. We have presented the initial mechanical and optical system-level tests that are currently underway in order to vet these new materials for use in CMB receivers. Provided our continued testing of prototypes is successful, we plan to deploy a science-grade multi-layer woven HMPE window on the \bicepthree~receiver in the coming year. We expect that replacing the current 1.25''-thick slab HDPE with a new multi-layer woven HMPE window will improve the sensitivity by 25\%.

\bibliographystyle{spiebib} 
\bibliography{report} 

\end{document}